\documentclass[sigconf]{acmart}
\AtBeginDocument{%
  }

\copyrightyear{2025}
\acmYear{2025}
\setcopyright{acmlicensed}\acmConference[FAccT '25]{The 2025 ACM
Conference on Fairness, Accountability, and Transparency}{June 23--26,
2025}{Athens, Greece}
\acmBooktitle{The 2025 ACM Conference on Fairness, Accountability, and
Transparency (FAccT '25), June 23--26, 2025, Athens, Greece}
\acmDOI{10.1145/3715275.3732202}
\acmISBN{979-8-4007-1482-5/2025/06}

\usepackage{subcaption}
\usepackage{csquotes}

\usepackage{subcaption}
\usepackage{csquotes}

\begin{document}

%%
%% The "title" command has an optional parameter,
%% allowing the author to define a "short title" to be used in page headers.
\title[You Cannot Sound Like GPT]{“You Cannot Sound Like GPT”: Signs of language discrimination and resistance in computer science publishing
}

%%
%% The "author" command and its associated commands are used to define
%% the authors and their affiliations.
%% Of note is the shared affiliation of the first two authors, and the
%% "authornote" and "authornotemark" commands
%% used to denote shared contribution to the research.
\author{Haley Lepp}
\email{hlepp@stanford.edu}
\orcid{0009-0003-9789-7415}
\orcid{0000-0002-0979-4009}
\affiliation{%
  \institution{Stanford University}
  \city{Stanford}
  \state{California}
  \country{USA}
}

\author{Daniel Scott Smith}
\email{danielscottsmith@stanford.edu}
\orcid{0000-0002-0979-4009}
\affiliation{%
  \institution{Stanford University}
  \city{Stanford}
  \state{California}
  \country{USA}
}
%%
%% By default, the full list of authors will be used in the page
%% headers. Often, this list is too long, and will overlap
%% other information printed in the page headers. This command allows
%% the author to define a more concise list
%% of authors' names for this purpose.
%\renewcommand{\shortauthors}{Lepp et al.}

%%
%% The abstract is a short summary of the work to be presented in the
%% article.
\begin{abstract}
LLMs have been celebrated for their potential to help multilingual scientists publish their research. Rather than interpret LLMs as a solution, we hypothesize their adoption can be an indicator of existing linguistic exclusion in scientific writing. Using the case study of ICLR, an influential, international computer science conference, we examine how peer reviewers critique writing clarity. Analyzing almost 80,000 peer reviews, we find significant bias against authors associated with institutions in countries where English is less widely spoken. We see only a muted shift in the expression of this bias after the introduction of ChatGPT in late 2022. To investigate this unexpectedly minor change, we conduct interviews with 14 conference participants from across five continents. Peer reviewers describe associating certain features of writing with people of certain language backgrounds, and such groups in turn with the quality of scientific work. While ChatGPT masks some signs of language background, reviewers explain that they now use ChatGPT “style” and non-linguistic features as indicators of author demographics. Authors, aware of this development, described the ongoing need to remove features which could expose their “non-native” status to reviewers. Our findings offer insight into the role of ChatGPT in the reproduction of scholarly language ideologies which conflate producers of ``good English" with producers of ``good science."
\end{abstract}

%%
%% The code below is generated by the tool at http://dl.acm.org/ccs.cfm.
%% Please copy and paste the code instead of the example below.
%%
\begin{CCSXML}
<ccs2012>
   <concept>
       <concept_id>10003456.10003457.10003580.10003584</concept_id>
       <concept_desc>Social and professional topics~Computing organizations</concept_desc>
       <concept_significance>300</concept_significance>
       </concept>
 </ccs2012>
\end{CCSXML}

\ccsdesc[300]{Social and professional topics~Computing organizations}

%%
%% Keywords. The author(s) should pick words that accurately describe
%% the work being presented. Separate the keywords with commas.
\keywords{Language ideologies, Peer review, ChatGPT, Discrimination, Semiotics}

%%
%% This command processes the author and affiliation and title
%% information and builds the first part of the formatted document.
\maketitle

\section{Introduction}
The first sentence of Bruno Latour's \textit{Science in Action} is an acknowledgment: ``Not being a native English speaker I had to rely heavily on my friends to revise successive drafts of this manuscript" \citeyearpar{latour1987science}. Through the following pages, Latour inscribes ---with a little help from his friends--- a method for examining the production of scientific facts and technical artifacts. Central to this method is the question of rhetoric: ``when someone utters a statement, what happens when the others believe it or don't believe it?" Statements believed are built upon, incorporated into other work, and make their way into the production of facts through collective, interactive labor  \cite{cheng2023new}. Disbelieved statements (and those who make them) are ``made more wrong" as they are left out of the network of scientific production. As Latour notes, clarity may be the first skirmish between the scientist and their reader in evaluating whether claims are believable. However, despite its ubiquity in evaluative guidelines\footnote{We analyzed 56 rubrics from the highest ranked scientific publications on \citet{scimago} related to the natural sciences and engineering. We classified all 504 criteria items in the rubrics into six common evaluative categories along which peer reviewers are instructed to evaluate scientific papers: motivation and impact, originality, correctness, thoroughness, replicability, meaningful comparison, and clarity. 83 (16\%) of the rubric items fall under the category of clarity. Other than attention to clarity, our classification categories align with the theoretical and empirical results of other scholars \citep{kuhn_objectivity_1977, lamont_how_2009, birhane2022values}, who have theorized variance in how scholars construct and prioritize evaluative criteria.}, clarity is regularly dismissed as a non-scientific evaluation criterion or delegated to other fields of study. For example, while the criteria through which statements are evaluated as trustworthy have been subject to scrutiny by sociologists, philosophers, and historians of science \cite{kuhn_objectivity_1977}, scholars of scientific peer review tend to focus primarily on epistemic values such as replicability or novelty \citep{lamont_how_2009, teplitskiy_sociology_2018, kennard2021disapere}. 

Evaluation of language clarity has particularly high stakes for researchers whose first language is not English. English has become so prevalent in scientific publishing as to be considered hegemonic. In computer science, for example, of the top indexed publications by Scimago in all languages, 100\% are published exclusively in English \citep{scimago}. This linguistic monopoly on indexed publishing has direct costs to individuals and social groups, from access to knowledge and knowledge production, to time spent in publishing, to career advancement \citep{amano2023manifold}. Consequently, automatic translation, and more recently, tools for English writing and speech generation, have been marketed to assist users in to acquiring cultural capital with currently-dominant language varieties \citep{ahmed2022app, warschauer2023affordances}. 

There is evidence that this use case is widespread and growing rapidly in academia. For example, \citet{naturemapping} has charted a rise in the use of ChatGPT by scholars for publishing and peer review. In January 2024, they estimate that approximately 17.5\% of sentences in computer science preprints uploaded to \textit{arXiv} were substantially altered by ChatGPT. While Liang et al. speculate that the faster rise in computer science could be related to higher computer scientist familiarity with the tool, this variation could also be related to the prestige of English-only computer science publication venues and the growing internationalization of the discipline. Moreover, the rate of growth of ChatGPT-generated sentences is higher for first authors affiliated with Chinese and European (non-U.K.) institutions in computer science preprints compared to authors affiliated with U.S. and U.K. institutions. As such, we center this study around the following questions.
\\
\textbf{RQ1:} Is the availability of ChatGPT associated with changes in writing evaluation for authors from different geographic regions?
\\
\textbf{RQ2:} How do perceptions about language appropriateness influence the way multilingual scientists engage with ChatGPT in computer science publishing? 

We examine the case study of International Conference on Learning Representations (ICLR), a large, influential computer science conference.  To answer RQ1, we analyze critiques and praise of writing ``clarity” across the conference's almost 80,000 peer reviews of over 20,000 papers before and after the introduction of ChatGPT. We interpret reviewer feedback on writing not as objective measurements of language but as expressions of language ideologies about what kind of writing should be published. To answer RQ2, we conduct semi-structured interviews with 13 language-marginalized peer reviewers and authors from across five continents and one area chair who has been with the conference since its founding. By focusing on the peer review process, we explore how a scientific community uses language to negotiate whose statements get believed, and whose are left out of the network of scientific production. 

\section{Signs of Clarity in Science}
Though there is increasing literature about how race, ethnicity, gender, disability, geographic, and other biases manifest in scientific peer review \cite{day2015big, helmer2017gender, iezzoni2018explicit, nielsen2021weak, murray2018gender, zilberstein2021national}, the nature of linguistic bias in scientific practice has received less empirical attention (exceptions include \cite{smith2023peer, canagarajah2002geopolitics, gordin2015scientific}). Moreover, though double-anonymous peer review has been shown to mitigate the effects of certain types of biases \cite{tomkins2017reviewer, sun2022does}, it does not mask language itself as a source of bias for peer reviewers \cite{herrera1999language}.

How is language evaluated in scientific writing? Formally, “clarity" is a criterion offered in peer review rubrics which can ambiguously refer, for example, to ways of writing, expressing ideas, or making arguments. Evaluation is conducted by the reader rather than the writer. As such, we focus our attention on reviewers. Like all people, peer reviewers express \textit{language ideologies} that link forms of language to ideas about social life \cite{woolard94}. As Gal and Irvine\cite{gal2019signs} argue, “statements about language are never merely statements. They entail ideological positions” which “have wide-ranging consequences in the material world.” Evaluations of clarity in peer review construct not just how language is best used in a scientific community, but the identity of who should be participating in science, and what sorts of information conveyed should be legitimated as believable and credible in the scientific canon.   

The role of language ideologies in the legitimation of social difference has long been documented in other settings. For example, Inoue \cite{inoue2004does} describes how male intellectuals in Japan during the 1990s used an imaginary of ``women's language" in premodern Japan to accuse contemporary schoolgirls of having ``lazy" speech; this contrast then was used to explain that contemporary women had been corrupted in modern Japanese society. Flores and Rosa \cite{flores2015undoing} offer an interpretation of language ideologies in the reconstruction of racial difference in U.S. schools. Through their concept of \textit{raciolinguistic ideologies}, they explain how, despite the fact that Standard English is something that cannot be empirically specified, ``non-standard" language varieties are routinely approached as deficits to be overcome among racialized students. Indeed, even when students change their languaging practices, they continue to be evaluated ``based not on what they actually do with language but, rather, how they are heard by the white listening subject." Applied to our case, this literature suggests that if we find language ideologies which link writing clarity and author demographics, we may not necessarily see that linkage break just because authors can use LLMs to adjust their writing. Rather, technical innovation may inspire social innovation: boundaries of distinction \cite{bourdieu2018distinction} between social groups may shift and new ways of marking otherness may emerge.

To test this prediction, we conceptualize the clarity evaluation process by drawing on the Peircean theory of signs, in which something (the sign) stands to somebody (an interpretant) for something (the object) in some respect \cite{peirce1974collected}. We are particularly interested in \textit{indexical} sign relations, through which there is a contextual connection between the sign and object; that is, the sign evokes something about the social world to the interpretant. For example, a scientist reviewing an anonymous manuscript comes across a plural noun with a missing \textit{s}. The absence of an \textit{s}, to the scientist, might index that the anonymous author of a missing \textit{s} could be from a different country. The interpretation may or may not be correct, but the fact that the reviewer associates the \textit{s} with author demographic, as well as how the reviewer makes sense of that association, has important implications for the peer review process. Directly, linking language to identity qualities could lead to differential peer review scores. Indirectly, even if such an effect were inconsistent, the perception by authors that such an effect is possible could lead them to adjust their writing or avoid submission altogether. 

However, there has been limited empirical analysis of how such an ideology might be reproduced in this setting, nor how language-marginalized scholars contest it. One exception is the literature about multilingual publishing. The proportion of scientific publications which are written in English has risen dramatically over the last century \citep{hamel_dominance_2007}. Today, 90\% of indexed journals (here, we refer to the homophonous indexing of scientific journals) in the natural sciences are published in English \citep{di2017publish} and the top 100 computer science publications according to Scimago are all published only in English \citep{scimago}. However, plenty of scientific publishing takes place in other languages. The challenge is that common journal indexers, including Web of Science and Scopus, leave out over 25,000 journals, 48.3\% of which are published in more than one language \citep{khanna2022recalibrating}. This publishing infrastructure may contribute to the growing number of citations from highly active ``core" countries and fewer of ``periphery" countries \citep{gomez2022leading}. Though some claim this infrastructure acts as a \textit{lingua franca}, in reality, access to English is uneven at best. Both between and within countries, the population who is trained to engage in academic English with ease is often of the upper class. As such, the global, monolingual publication industry contributes to a lack of diversity on other dimensions, such as region and socioeconomic status, which in turn contribute to the oft-cited ramifications of low diversity in academia, from low scholarly attention to specific issues \cite{septiandri2023weird, ranathunga2022some}, to perceptions that English is the (only) language of technology \cite{leblebici2024you}, to diminished scientific legitimacy for much of the world's knowledge practices \cite{wynter2003unsettling}. As a result, many scholars will pursue English-language scientific publishing. How they navigate this pursuit, and how they are treated, is the subject of this paper.

By combining an analysis of peer reviews from ICLR over the last seven years with interviews of computer scientists around the world, we offer empirical evidence for how an English-only publishing community polices its linguistic boundaries. We expand on the existing theoretical literature by examining how language technologies come to mediate the expression of language ideologies, and by attending to the role that language ideologies play in the construction of scientific knowledge.

\section{Empirical Setting: The Case of ICLR}
ICLR is a highly ranked, peer-reviewed computer science conference attended by thousands of academic and industry researchers from around the world. Publishing in this conference is consequential; computer science conference papers are comparable in prestige to journal articles in other disciplines. We examine discussions of writing clarity in ICLR peer review in the period starting from its first double-anonymous conference in 2018. There are several reasons this venue is a useful case study for our research questions. 

First, the challenges of English-only publishing are particularly concentrated in the discipline of computer science. In computer science, publishing in one of the global top venues requires that findings must be described in English. As such, English-language venues represent a gateway through which computer scientists around the world must pass to participate in the global disciplinary community. ICLR is an internationally diverse conference with authors from all over the world, so reviews evaluate writing of authors from many different language backgrounds. 

Second, the rise of LLM-use in this community offers a mechanism for comparing how language ideologies can be expressed when writers have new ways to mask their language of origin. \citet{naturemapping} find that rates of ChatGPT-modified sentences in preprints started rising faster in computer science than in other fields. At the time of writing, there are only three years of conference proceedings since the release of ChatGPT, so studying the field which demonstrates earlier adoption provides more data as well as an indicator of what might occur in fields that could see later adoption. However, as preprints do not have peer reviews, \textit{arXiv} preprints offer limited signals as to how readers react to changes in text. ICLR's publicly-published peer reviews give rich examples of language ideologies expressed by reviewers and reacted to by scholars. 

Furthermore, ICLR's 2018 iteration was the first full conference cycle after the influential invention of transformers, an invention which led to a massive expansion in AI research. After this invention, funding for scholarship ballooned \citep{ahmed2023growing} and large numbers of authors from across academia and industry started submitting to conferences such as ICLR. Though computer science as a discipline has long informed innovation in science and industry, the increased valuation of AI research in this period increased the influence of ICLR as a conference not just on the research community, but on innovations which have been rapidly adopted by organizations and individuals around the world \citep{noauthor_iclr_2020}.

\begin{figure}
\raisebox{-\height}{\includegraphics[width=\columnwidth]{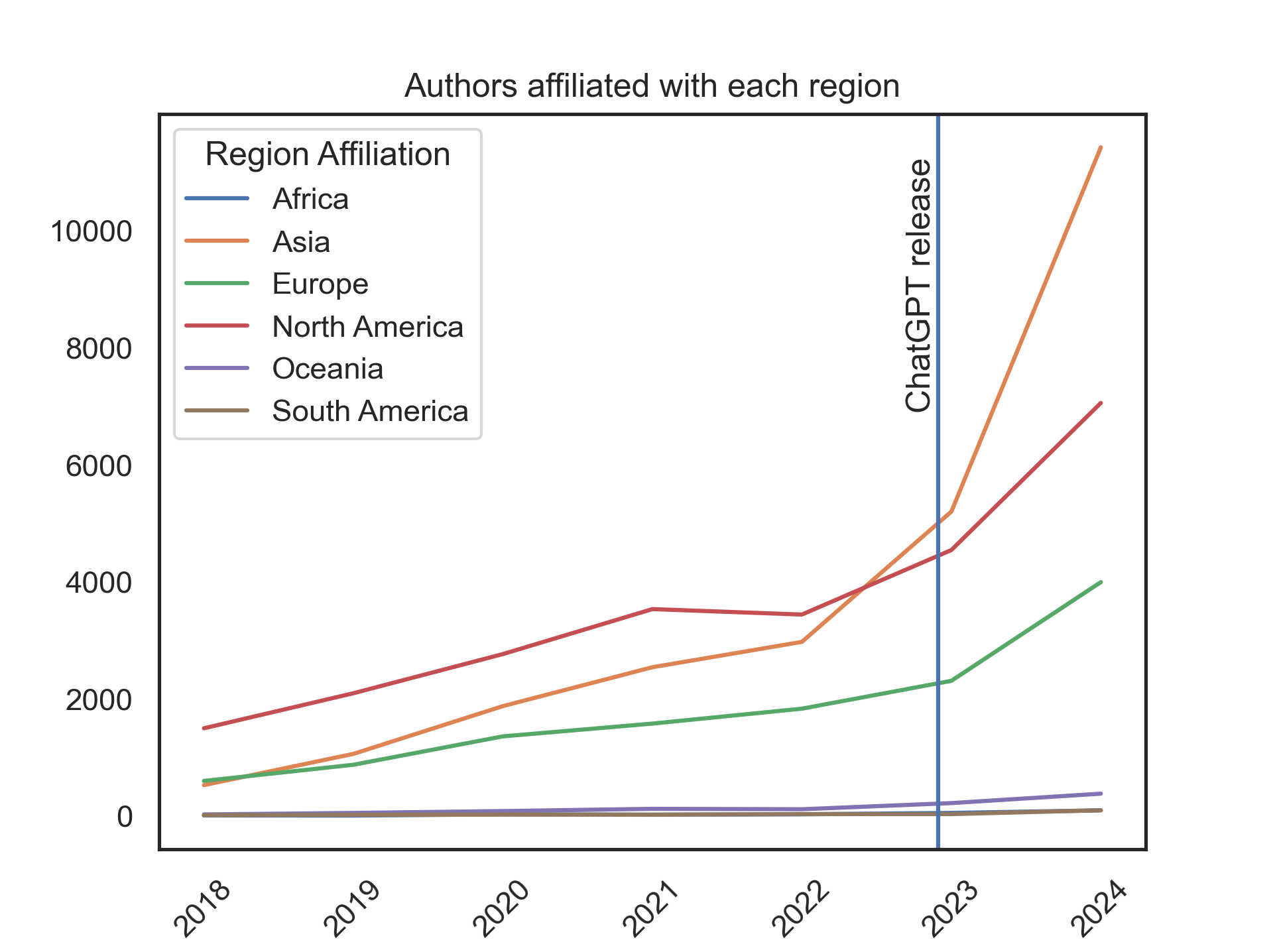}}
\caption{ICLR has seen a rapid increase in submissions and author internationalization. The vertical line marks the release of ChatGPT on November 30, 2022.}
\Description[Line graph of author internationalization over time by region] {Line graph of author internationalization over time showing rise of all authors, with greatest rise from Asia}
\label{fig:international}
\end{figure}  
\begin{figure}
    \raisebox{-\height}{\includegraphics[width=\columnwidth]{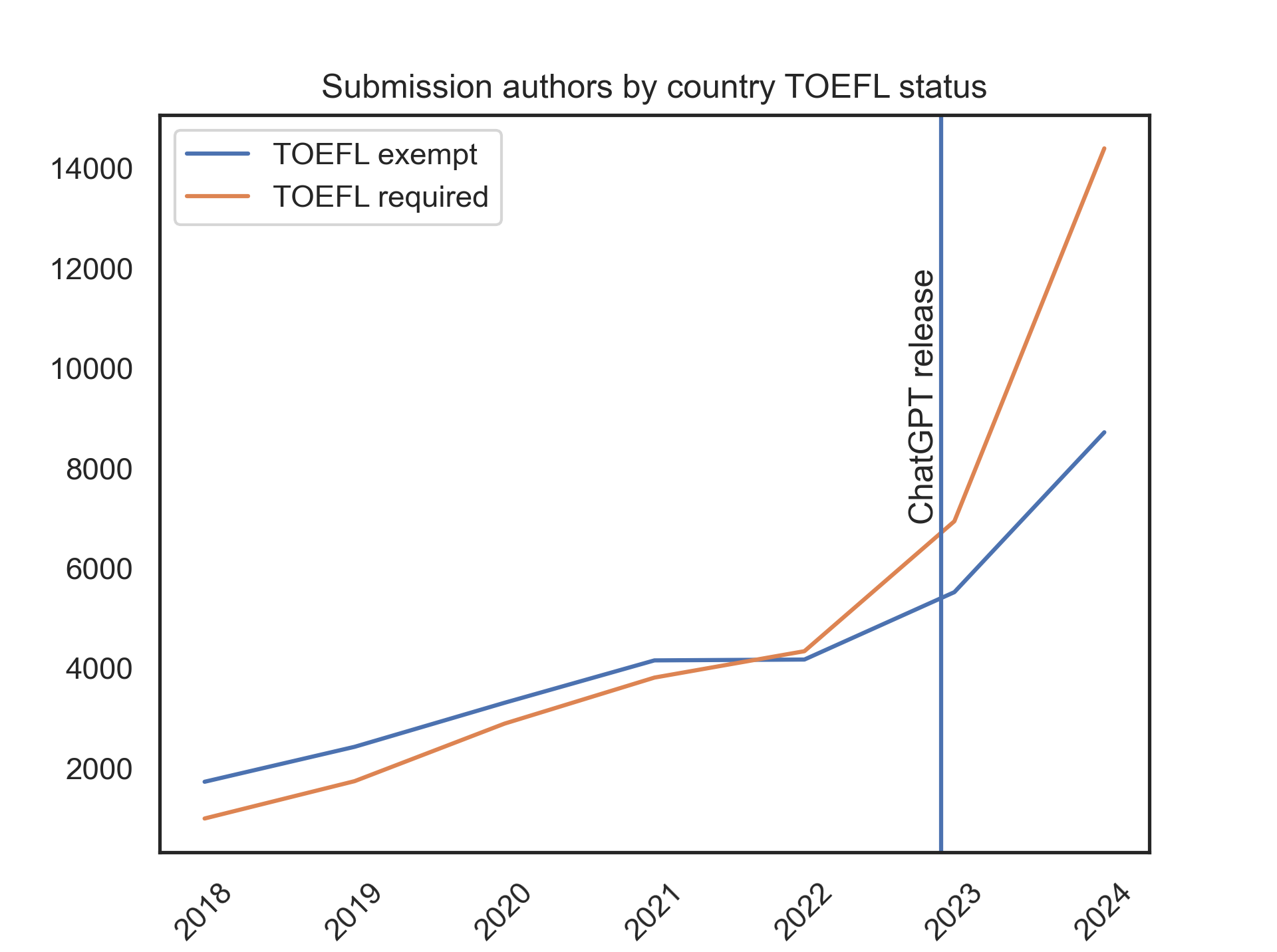}}
    \caption{The number of authors from countries which would require a TOEFL score for admission at a U.S. university surpasses authors from TOEFL-exempt countries between 2022 and 2023.}
    \Description[Line graph of author internationalization over time by TOEFL cateogry] {Line graph of author internationalization over time showing rise of all authors, with greatest rise from TOEFL-required countries}
\end{figure}

\section{Methods}
In this section, we outline the methodological details for our statistical study of peer reviews and our interview study of authors and reviewers. Our quantitative work relies on publicly available data; our qualitative work is approved by the Institutional
Review Board (IRB) Protocol 74594 at Stanford University.

\subsection{Statistical analysis of regional differences in writing evaluation}
To answer \textbf{RQ1}, we draw on the complete set of peer reviews from ICLR 2018-2024 to measure the association between reviewer critiques of writing and the world region of authors. We then describe how that regional differentiation shifts after ChatGPT became available. 

\subsubsection{Data}
ICLR publishes all reviews and conversations between authors and reviewers for all accepted or rejected submissions. Until the database for each year's conference is published, author information is not shown to reviewers and reviewer information is not shown to authors. Hypothetically, this means that authors and reviewers do not know the specific identity of their respective conversants. Authors and reviewers may guess at elements of an anonymous interlocuter's identity based on language traits or preprints, which we discuss in Section~\ref{section:qual}. Using the API of OpenReview.net, we extracted 76,453 reviews of 20,827 posters, spotlights, and oral presentations submitted to ICLR between 2018 and 2024. We include only the texts which come from reviewers. 

We analyze three dependent variables as described in Appendix ~\ref{appendix:variables}: the standardized review score given by a reviewer (higher is better), the standardized number of sentences in a review in which a reviewer critiques writing (lower is better), and the standardized number of sentences in a review in which a reviewer praises writing (higher is better).

Our main independent variables include indicators of where authors write from. We label the country of the author's earliest listed institution in OpenReview (Appendix \ref{appendix:variables}). While country may act as a proxy for language background, it is a deeply inexact match. Therefore, we sort country data in two ways: by continental region, and by institutionalization of English as defined by TOEFL score\footnote{The Test of English as a Foreign Language (TOEFL) is used by 11,000 universities and other institutions in over 190 countries and territories.} exemption policy of Stanford University, which mirrors the policies of universities around the world in which English is the language of instruction. The TOEFL exemption policy therefore represents the way that many institutions act on the relationship between language and country of origin in practice.  Since our dependent variables describe manuscripts, not authors, we aggregate these indicators as the percentage of authors writing from each place. 

In the period we examine, ICLR experienced substantial demographic change as many more scholars from Asia began to participate in the conference. Based on the number of scholars represented from each region (Figure \ref{fig:international}) and  speculative discussion by interviewees about Chinese scholars (Section \ref{ref:qualityscience}), we report results for scholars associated with Asian and Chinese institutions specifically. 

Our next set of independent variables include indicators for time, dichotomously measured as pre-/post-GPT (November 2022). This allows us to split the corpus of reviews  respectively and then descriptively explore changes in evaluation patterns associated with the availability of GPT.

For controls, we include reviewers' other epistemic value judgments (Appendix \ref{appendix:variables}). Like with critiques and praises of clarity, we infer these at the sentence level using training data \citep{hua2019argument} to fine-tune two RoBERTa  models \citep{devlin2018bert}: one for binary classification of polarity and then a second, separate one for classification of six aspects or criteria of merit. We combine polarity and criteria labels at the sentence level and then sum the number of sentences containing the resulting value judgments in each review. We standardize these sums across the corpus (Appendix \ref{appendix:variables}). Finally, we include controls for review length and for the general internationalization of the author pool in a given conference year with the Shannon Diversity Index.

\subsection{Statistical Approach}
We fit several panel OLS models with manuscript random effects of the general form:
\[
Y_{mr} = \alpha + \boldsymbol{\beta} \cdot \mathbf{X}_{m} + \boldsymbol{\gamma} \cdot \mathbf{Z}_{mr} + u_{m} + \epsilon_{mr}
\]

where, for reviewer \textit{r} of manuscript \textit{m}:

\textit{Y} is one of three dependent variables, including: the standardized number of sentences a reviewer negatively evaluates writing clarity (`Clarity (–)'); the standardized number of sentences a reviewer positively evaluates writing clarity (`Clarity (+)'); or the reviewer's standardized rating (`Rating').

$\alpha$ is the y-intercept, representing the mean of the respective dependent variable when all independent variables are zero.

\textbf{X} is a vector representing our independent variables and controls that describe the manuscript but do not vary among reviewers, including: the percentage of authors associated with institutions in TOEFL-required countries, Asian countries, or China; and the Shannon Diversity Index value describing the degree of internationalization among authors in the conference year the manuscript was submitted (a higher value indicates greater diversity among submitting authors, while a value closer to zero indicates lower diversity);

\textbf{Z} is a vector representing our independent variables that describe the review of a manuscript and so vary among reviewers, including: their epistemic value judgments, such as those pertaining to accuracy, novelty, consistency, thoroughness, and replicability and the length of their review.

\textit{u} is the random effect associated with the manuscript, which accounts for differences between manuscripts and is needed because the nesting of reviews within manuscripts violates the assumptions of ordinary least squares;

$\epsilon$ is the idiosyncratic error term, which captures measurement error and other influences on \textit{Y}. 

We split our corpus into pre- and post-GPT periods and run our models separately on each. This allows us to account for changes in time in general and to \textit{descriptively} explore whether the evaluation patterns in ICLR shifted after GPT was introduced (i.e., whether the strength in the association between author background and clarity critiques changed between the two periods). As such, we interact all our independent variables with the pre-/post-GPT indicator. Any differences in estimated coefficients between the two periods suggest correlational evidence that evaluation patterns changed after the introduction of GPT.\footnote{Our code is available at \url{https://github.com/hlepp/sound_like_gpt}.}
% Example quotes
\begin{figure*}
\fbox{
    \begin{subfigure}[t]{0.45\textwidth}
         \centering
``The authors are either novices, not native English speakers, or the PI didn’t bother to help writing it." 
    \end{subfigure}
\qquad

    \begin{subfigure}[t]{0.45\textwidth}
         \centering
``The paper is full of English mistakes. A proficient English speaker should correct them."
    \end{subfigure}}
\caption{Sentences from reviews which our method infers are about clarity also make inferences about the identity of authors, echoing our findings in qualitative interviews.}
\Description[Two quotations from peer reviews.] {``The authors are either novices, not native English speakers, or the PI didn’t bother to help writing it." ``The paper is full of English mistakes. A proficient English speaker should correct them."}
\label{fig:example_labels_b}
\end{figure*}

\subsection{Interviews with multilingual conference participants}
\label{section:qual}
To answer \textbf{RQ2}, we conducted interviews with multilingual scholars who participated in ICLR in 2023 or 2024, and one area chair (AC) who has been involved with ICLR since its founding.
\subsubsection{Participant Recruitment}
Data about the language identities of scholars are not regularly collected or verified by publishers, so it is not possible to conduct a random sample of this population. Instead, we use purposive snowball sampling of scholars who submitted to ICLR after the release of ChatGPT and whose first language is not English. Between November 2023 and August 2024, we advertised extensively using computer science-related Slack workspaces and the social media platform X. The recruitment messages, which were written in English, were shared by several prominent scholars in the ICLR community, within the ICLR Whova application, and by the official ICLR X account. The X post was shared widely, with analytics indicating 23,600 views; 24 retweets, 48 likes, 13 stars. We also sent email invitations to a random sample of ACs who had worked with ICLR from 2021-2022. We received responses from 25 interested scholars, of which we eventually interviewed 14. We can only speculate about why we struggled to recruit more interviewees, but others (e.g. \cite{ali2023walking}) have noted the unique difficulties of recruiting technology workers due to concerns about non-disclosure agreements (NDAs), privacy, and discouragement from company legal teams.

Because of our recruitment requirements, our sample was highly educated with all interviewees either having or pursuing a graduate degree. We intentionally recruited scholars with who identified as having a first language other than English in order to understand the experience of language-minoritized scholars. In addition to English, interviewees described speaking many languages, including Arabic, Cantonese, Fon, French, German, Gu, Hokkien, Italian, Kabyle, Mandarin Chinese, Portuguese, Russian, Spanish, and Swiss German. Out of the 13 language-marginalized interviewees, ten were currently residing outside of the countries they grew up in. All of the ten had left their home countries in Africa, East Asia, and South America and were currently working or studying in the United States, United Kingdom, Canada, and Singapore. Of the three who were currently residing in their home countries, one had previously been a visiting scholar in another country, and one was seeking opportunities abroad. As such, many interviewees reflected on their experiences not just with publishing in English, but in using the language to navigate the social, cultural, and financial difficulties of academic immigration. We do not report AC regional data to preserve confidentiality. All interviewees had experience peer reviewing, and most had been peer reviewers for ICLR during this period.  

The first author conducted 14 semi-structured interviews. Her first language is English and she is affiliated with a U.S. institution. She holds a graduate degree in Computational Linguistics and worked in Natural Language Processing research for several years. She prepared for the interviews by researching the history of ICLR and similar conferences, reviewing the publishing requirements of the 56 top engineering and natural science publications, conducting ethnographic observations of academic computer scientists, and participating in an ICLR workshop. 

During interviews, we informed participants that we were interested in learning about the use of AI in scientific publishing and in understanding the experiences of diverse scholars in the computer science community. Because we could not compensate interviewees and were interviewing busy scholars during conference season, interviews were often bounded inside 30 minute appointments (though we frequently went well beyond that as people described the interviews as ``interesting" and ``meaningful").  

We followed institutional standards for harm reduction associated with our IRB. We took particular care due to the common stories of immigration of our interviewees \cite{bloemraad2022precarious}, and report results with minimal demographic information to preserve confidentiality. Our IRB has different restrictions on data collection for people in different countries, with the strictest requirements related to interviewing people who were physically located in China during the time of the interview. In this case, we took detailed notes instead of recording. For all interviewees outside of China, we asked for oral consent to interview and to record. The first author wrote a memo after each interview about initial impressions from the talk, then listened to the interviews again to manually ensure automatic transcriptions are as faithful as possible to the conversation. We do not list names or references to institutions to preserve anonymity.  We collected 6 hours and 43 minutes of footage plus approximately 2 hours from 2 interviews not recorded for IRB reasons; 34 pages of notes; and 229 pages of transcripts.

Using abductive analysis \cite{timmermans2012theory}, we coded each interview in NVivo. After the emergence of ``anomalous" and ``surprising" themes around indexicality, we re-coded the data to examine the indexical relationships described. We created 30 codes: 23 related to general language labor and 7 related to indexicality. Interviewees would often switch between discussing their experiences in their capacity as author and in their capacity as reviewer or AC; we attend to these differences in our coding. Additionally, many scholars, cognizant of changes happening within their discipline, would speculate about trends in the field as a whole. We were careful to differentiate between discussions of field-level speculation and descriptions of personal experiences. We emphasize that we cannot, nor do we want to claim that our interviewees describe generalizable experiences. Rather, our results show what can happen, and indeed does happen, to language-marginalized scholars in international computer science publishing.

\begin{figure*}
        \raisebox{-\height}{\includegraphics[width=\textwidth]{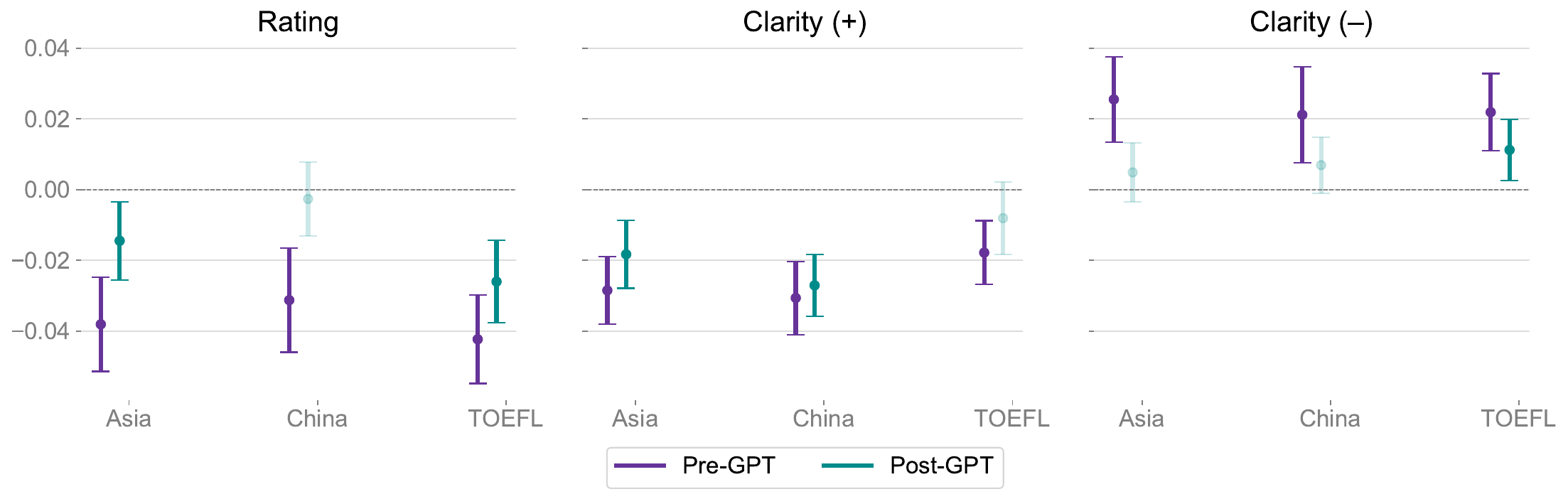}}
        \caption{Shifts in reviewer scores on overall paper ratings, praise of writing clarity, and critiques of writing clarity before and after ChatGPT becomes available in 2022.}
        \Description[Error plot showing changes in paper ratings before and after GPT] {Error plot showing lower review scores and clarity scores for authors from Asia, China, and TOEFL req countries, slightly ameliorated after GPT released.}
    \label{fig:stat_results}
\end{figure*}
\section{Findings}
\subsection{Clear bias but unclear solutions}
\label{ref:clearbias}
We summarize our inferential modeling in Figure \ref{fig:stat_results} above. Each x-label corresponds to the indicator of author place, measured as the percentage of authors on a manuscript. The y-axis measures the estimated coefficient on each indicated dependent variable measured from the review of that manuscript. Our findings are consistent with other work suggesting that multilingual scholars experience what others have called a ``tax" in publishing \cite{amano_manifold_2023}. Across all  outcomes, papers with a higher percentage of multilingual authors fare worse: they receive lower ratings; fewer praises on clarity; and more critiques on clarity, even after statistically accounting for the other substantive content of the reviews. We also observe suggestive but inconclusive evidence that the availability of ChatGPT may have reduced the expression of this bias (all else in the model equal). This is evident in the general vertical shift in the estimates toward zero between pre-ChatGPT and post-ChatGPT periods. Yet we note that for most demographics, these shifts are not statistically significant (i.e., the error bars overlap between pre/post periods). The exception is scholars in China, who appear to be rated no more harshly than the general population after ChatGPT becomes available.

Based on the attention given to generative AI models as salvos against language discrimination, we were surprised by the muted improvements to equity of clarity critique and praise. However, our tools are limited: our evaluation is not causal, so we can only describe the shifts over time rather than attribute them to a particular intervention or authors' specific usage of these models. 

Such macro-level descriptive trends motivate deeper inquiry and an examination of \textit{science in action} \cite{latour1987science}: how do scientists interpret and engage with linguistic difference? How do they make sense of generated text? To address these questions and to get a better understanding of the scholars the trends above represent, we turn to our interview data, which offers several explanations for the continued presence of regional inequality of writing evaluation in computer science publishing.

\subsection{Language labor}
\label{ref:languagelabor}
Throughout our interviews, we heard examples of the substantial labor that language-marginalized scholars engage in to participate in the computer science research community. This labor is largely hidden in our statistical data because much of it happens long before the writing of a paper. From decades of English education, to global migration, to physical exhaustion, interviewees put substantial time, money, and social effort into reading, writing, speaking, and listening in English. Though most interviewees described the use of ChatGPT in the very last stages of the writing process (``4 days before submission, panicking!"),\footnote{Liang et al.\cite{naturemapping} find that number of ChatGPT-generated sentences is higher for papers submitted right before a deadline.} interviewees discussed language labor in their work and life long before a paper was prepared for publication.

For example, many interviewees described the importance of their social networks as they adjusted to new linguistic environments. Yet when reflecting on the changes to their experience since ChatGPT, several described a decrease in community support. 
\begin{displayquote}
At that time [before ChatGPT], like the senior students in our lab and the supervisor will help me somehow, or change my writing more instead of now... Now they somehow will try to only give this high level [feedback] and ask me myself to do this kind of paraphrase, or grammar check, or something. Previously they will give more efforts in that part.
\end{displayquote}
Nobody described the tool facilitating more collaboration across language backgrounds; on the contrary, many scholars, especially those raised in East Asia, described building homophilous networks by language background: ``what's happening is that I have more collaborators in Chinese... I think maybe one of the reasons is that since, I'm Chinese, so I have many Chinese friends... so maybe this connection will make it easier to collaborate." This homophily was rarely described as a choice. Several people mentioned they hoped to work with ``native speakers" some day. At the same time, people described writing about computer science in other languages than English to be convoluted,  difficult to read, or even laughable. One expressed the difficulty of attending computer science lectures in languages other than English. Rather, scholars shared the experience of existing in the liminal linguistic space of not conducting research in a first language, and thus engaging in extensive mental labor to produce work in a way that would be as close as possible to ``native speakers." ``Native speaker" was not a status any of the interviewees described hoping to achieve.

We describe the reasons interviewees gave for participating in English-language computer science publishing, because these reasons begin to explain the linguistic ideologies we observe in the peer review process. Though the affordances of ChatGPT would support translation not just into English, but out of English into other languages, no interviewee was using it for this purpose. In fact, when prompted, interviewees laughed off the idea. Most of our interviewees explicitly described associating non-English computer science publications with lower quality research. One scholar admitted 
\begin{displayquote}
I have a bias myself... if the research is written in a language that is not English, I assume that it's of lower quality. I'm sorry to put it this bluntly, but, like. Yeah. 
\end{displayquote}
Concerned with the implications of this lower quality, from impact to readership, this person decided to avoid publishing in their first language:
\begin{displayquote}
I did one research project in Spanish. But then, when we actually wanted to publish…  we submitted to <a major AI conference>. Of course that was in English, and I don't think anyone has ever read, the [Spanish one], not even my advisors.
\end{displayquote}
A second scholar explained that their choice of English was related to resources: 
\begin{displayquote} Honestly, I'm biased toward English... there are way more resources in English, you know, and a lot of those technologies I define are created in English and work in the programming and everything. Everything is in English, you know.
\end{displayquote}
Though this person tried to attend a lecture in their first language, 
\begin{displayquote}
I was lost and I went directly back to English like, I couldn't just synchronize it. A lot of those technologies are just way better and well-described in English.
\end{displayquote}
A third person explained an association between English publishing and prestige:  
\begin{displayquote}Especially in computer science... I think we only focus on those kind of top conference papers. And yeah, none of them are non-English.
\end{displayquote}
This person and others explained that they saw learning English as part of the education required to become a computer science scholar. In short- rather than the affordances of the ChatGPT guiding adoption, it was visions of of how science works currently -that is, in English- that seemed to be at the wheel.

\subsection{Signs of Quality Science}
\label{ref:qualityscience}
A narrative of indexical linkages emerged from our interviews. In both their capacities as reviewers and authors, interviewees described how certain writing features index an author’s geographic or linguistic identity. The specific features of writing that scholars describe as signs that an author’s status was ``non-native” were typically grammatical or morphological. As one interviewee described,
\begin{displayquote}
    the present tense or… like leaf or leaves so…How to say this? Let me let me check about the accurate [word]. <checks translation> singular or plural…Yeah. So if I come across a lot of typos and grammar errors, I would say, I would guess like this, this must [be a] non native speaker [or] at least is not from U.S… or England.
\end{displayquote}
Others also listed tense and plural use, as well as long ``clunky" sentences as being indicators of language identity. Notably, tense and plural deletion and long sentences are documented as typical for Mandarin speakers of English, as Mandarin has serial verb constructions and does not have tense or plural markers. 
Several interviewees also described how these language forms index the level of quality or trustworthiness of the science described. As one scholar put it, 
\begin{displayquote}
If the English is not, it's not very legible, you start thinking like, oh, if they didn't put effort in passing this through Grammarly, yeah, you know, how does that reflect on everything else? And maybe that's maybe that's some discrimination that we shouldn't be doing like, not the correlation that we should necessarily be doing.
\end{displayquote}
The interviewee’s explanation of writing indexing writer attributes was part of a pattern we heard across our interviews: to our interviewees, grammatical errors evoked an author’s language background, and an author’s language background evoked the quality of the manuscript. In this chain of signs, the author’s identity appears to mediate the indexical linkage between the writing and the quality or trustworthiness of science. 

The way interviewees interpreted the linkage between author identity and writing quality varied. It was notable that many of the linkages peer reviewers described were between perceptions that authors were from China, and perceptions of how being from China influenced science quality. One interviewee described how
\begin{displayquote}
    There's now way more reviewers that are from non-western countries, in particular from China, and I'm not sure exactly what the consequences of that are. But I think you know, probably a consequence is that basically you know, less desire for incisive critical thinking.
\end{displayquote} 
This person described indexing ``less literary" writing with Chinese scholars, which he then indexed with a style of science that was ``more competitive and more conformous [sic].”  Another scholar, who himself was from China, described that when he detects papers that he believes are written by authors in China, he tends to grade them more harshly, because a certain level of English is required to study computer science at elite universities. When prompted to ask how he knew authors came from China, the scholar described certain phrases, and explained that he himself tried to avoid these phrases so reviewers would not detect his country of origin.

Indeed, many of the language-marginalized scholars we interviewed anticipated that reviewers might link country status to science quality, and described trying to avoid that association by changing their writing. When asked whether they thought reviewers valued international diversity, one scholar replied: ``What’s the opposite of value? From a prestige direction, I think many ICLR reviewers would prefer their reviews to be less internationally representative.” They also speculated that based on the connection of certain grammatical features with author country identity, reviewers might respond differently to errors: 
\begin{displayquote}
I feel like there’s a very entangled relationship between language and national identity. And so like any biases that might exist around language, it may not be about the language itself so much about the corresponding national identities… I think there’s a relationship between how lenient reviewers are with linguistic stuff and how much prestige they associate with the source of the paper. Like if you read one paper, and it has several typos, and it comes from Princeton vs. a school in a country where the majority language isn’t English. People will interpret that differently. The first one is attributed “oh this person was in a rush” and the second one to “lower English ability."
\end{displayquote}	

Two other interviewees explained that in their experience, the content of their scholarship was indexed by reviewers as low value by virtue of it not being \textit{about} English. As one shared, 
\begin{displayquote}
    I sort of had the impression that when I submitted work like this [in another language] that it was a bit hard to convince reviewers of the value. I guess it could have been easier if if the dataset was in English or in the U.S.
\end{displayquote}
In short, rather a direct indexical linkage between writing features and manuscript readiness, our interviewees describe a much more complex web of indexicals linking writing features to certain groups of people, and certain people to the quality and value of science for publication. 

\subsection{Limited Signs of Change}
\label{ref:signschange}
After the release of GPT, interviewees described how ``grammatical errors" and other signs indexing author language background began to disappear.  Instead, interviewees detected new linguistic cues that an author’s first language was not English, and continued to index the author’s language background with quality of science. In their capacity as reviewer, one interviewee postulated: 
\begin{displayquote}
Before, ChatGPT, I would say, I, I always come across paper[s] that are like, have a lot of grammar errors... they have very long sentences such like that it's very hard to understand the whole paper. But after ChatGPT, I would say maybe some author[s], they use that [ChatGPT] to polish or correct a grammar error. Then yeah, the writing at least, it's grammatically correct.
\end{displayquote}
A second interviewee, also speaking as a reviewer, described certain writing features indexing ChatGPT use, and ChatGPT use indexing writer language background:
\begin{displayquote}
If I find this paper to be well, majority like, kind of mostly written in GPT-style, then it also has a negative impact. So it would leave me a negative impression on this. I think this [goes] back to the same question, like that usually non-native speaker[s] tend to use this more. And also they tend to like, not spot [idiosyncrasies] in these kind of phrases that are written by ChatGPT, but also not common in English papers.
\end{displayquote}
A third reviewer explained how AI features indexed a person's ability to write correctly, which they indexed with trustworthiness:
\begin{displayquote}
    I think I have to admit that this [features of AI] will, to some extent have an active impact on my review, and I think the reasons is also pretty straightforward. Is that like, if you can’t make your writing right, then how can I trust you?...I know it is kind of unfair. But I can also have this impression.
\end{displayquote}
The interviewee paused and cringe-laughed, recalling that she had just described how ``errors" in writing before ChatGPT also limited her trust in multilingual authors: ``It's damned if you do! Damned if you don’t!"

Interviewees explained that words like ``delve," as well as flowery language indexed ChatGPT use, and through that use, of ``non-native" writing.  On the other hand, certain writing features indexed that ChatGPT was not used, and that lack of use indexed that a speaker's first language is English. 
\begin{displayquote} 
For some paper[s] they use a lot of jargons or, yeah, it's not that AI style, [so] I would guess, yes, that [it] must from some native speaker writings. Yeah. I would say, there's there's some latent AI style [that] I could guess, like sometimes it repeated a concept again and again. And sometimes [if] I saw a paper, they are very concise and accurate, and and they use some jargon, [so] then I would say, that's from some expert in this field, or at least they are very experienced.
\end{displayquote}

Speaking as a reviewer, the interviewee takes the cues from writing that indicate it is not ChatGPT-generated, and from there infers that the author must not just be any person, but a ``native speaker." The ``AI style” indexed an author whose first language is not English, and the absence of that style thus indexed an author whose first language is English. To them, writing that is ``concise and accurate" indexed not just a first language English speaker, but an expert or ``very experienced" person. 
Another interviewee described how the indexicality of writing errors motivated their engagement during rebuttals to peer review: 
\begin{displayquote}
    When I am receiving reviews and I see that some review is in super broken English, I think I can fight with the AC better. You know? So it's like… it's a signal that the reviewer didn't put that much effort. So I can, you know, I feel like I have a better chance of making the AC ignore that reviewer... [These days] if it sounds like ChatGPT, I feel like… I haven't had to have that fight yet. I've been very lucky. But the similar thing could be the case like, if it's if a review sounds like a generic thing generated by GPT, you can probably fight it better.
\end{displayquote}

Language-marginalized scholars described how the perception that other reviewers would act in this way informed their own use of the tool in scientific writing. While discussion of fixing grammatical errors typically centered around reaching a wider audience and gaining the legitimacy that comes with grammatical correctness, scholars using ChatGPT described using the tool in a way that would hide features of writing they think will lead to them receiving the label of ``non-native." One explained that``you cannot sound like GPT." Another person speculated that use of the tool was widespread, and perceived that such use would become evident by virtue of the fact that it was widespread:
\begin{displayquote}
Since we [people who are not first-language English speakers] use AI a lot, I'm kind of worrying. Like in the future, if we do not have our own style, we will all follow, like, AI style. That would not be very good for the like whole society, or like for the development of English... So I'm worrying like, as this [AI] centralizes, centralizing again [and] again, then some variety, some like style or features that are not very popular will be destroyed.
\end{displayquote}
To this scholar, the style (sometimes described by interviewees as ``the voice") of ChatGPT replaced the variation that people from different language backgrounds bring to their writing, and would continue to mark them as different in the long run.

\section{Discussion}
When we started conducting interviews, we sought to understand how scientists connected features of writing with evaluations about whether that writing was publishable. To do so, we probed interviewees to understand the features of writing that indicated to them that a manuscript was (in)appropriate for publication. However, as discussed in Section \ref{ref:qualityscience}, interviewees would often discuss not the qualities of the manuscript or its legibility, but the identity of an imagined author or group of authors. For example, most interviewees described how grammatical errors, as assessed by the readers, evoked the presence a ``non-native English speaker.” Notably, the imagined author was typically framed as a ``speaker” rather than writer; the grammatical errors tied not just to writing skill, but the imagined person’s generalized language identity. 

Interviewees described how imagined author(s) indexed the quality of science submitted. Though indexes, on their own, only suggest that there is a contingent or co-occurring relation between sign and signified, their social power comes from  ``\textit{ad hoc} hypotheses” made about that relation \cite{keane2003semiotics}. We begin to see these hypotheses as interviewees described that the presumed presence of a multilingual scholar could indicate that science was untrustworthy or low-quality. Alternatively, one reviewer suggested that an imagined Chinese author group could indicate that the science ``from several points of view, much higher quality” but the result is ``less room for, you know, real originality.” 

Though highly ranked computer science publishing today takes place in English, the way that people interpret and make sense of this fact is hypothetical. Interpretations are tools for efficiency \cite{merrell2005charles}. The hypothesis that an author's first language is not English, so therefore their science will be bad, is a way to quickly make an evaluation rather than examine the work of a specific person in a specific case. In this way, language status can act as a proxy for science quality, creating what Tilly \cite{tilly1998durable} calls a ``durable inequality" that stabilizes differences in resources and power into a ``status difference" between ``types" of people \cite{ridgeway2014status}. We hypothesize that the fast-paced nature of peer review in computer science at the time of writing may contribute to this phenomenon. Though it is outside the scope of the paper to conduct archaeological analysis of the associations which our interviewees describe, we can speculate two origins.

First, the descriptive words given by interviewees closely echoed findings of research on anti-Chinese sentiment in educational contexts, for example, that Chinese students lack critical thinking skills, are prone to plagiarism, and generally harm the educational environment \cite{moosavi2022myth}. Recently, there have been discussions about the presence and influence of these tropes in the AI community \cite{ha_neurips_2024}. The U.S. government has used similar language in letters to academic institutions (including our own) about Chinese AI scholars \cite{Levinletter}, and has explicitly targeted Chinese academics with migration restrictions, increased surveillance, and other forms of hostility \cite{Lu2025}.

As our quantitative data indicated and several interviewees explained, there is indeed a rise in Chinese scholars submitting to ICLR, many of whom are first time submitters. While our indicator of author diversity in the conference year shows mixed relationships with individual reviewers' evaluation patterns, the concept of indexing helps us to theorize the process through which the perception of demographic change can influence how individuals interpret difference. 

Secondly, the association of low-quality science with English-speakingness may be related to resources, impact, and prestige of English-language scientific publishing discussed in Section \ref{ref:languagelabor}. Scholars draw on their experiences and knowledge with the way science functions today, and that knowledge may inform how they come to index writing features with people, and people with science quality. As Keane describes, ``signs give rise to new signs, in an unending process of signification," and signification can, to the interpreter, ``entail sociability, struggle, historicity, and contingency." \cite{keane2003semiotics}.

Changes in this indexical chain demonstrate how markers of social status may shift over time \cite{elias1939civilizing}. When ChatGPT became widely used in scientific writing \cite{liang2024monitoring}, and the linguistic features that interviewees associated with an author’s language background began to disappear, interviewees described that they found new indexes for people from different language backgrounds. Even though the writing was now ``grammatically correct,” interviewees described an indexical connection between certain words (e.g. ``delve” or ``meticulous”) and a text being generated by an LLM, as well as a connection between the LLM being used and the user being a person with a certain language background. This result parallels Flores and Rosa's \cite{flores2015undoing} description that even after U.S. students adapt their speaking style for a \textit{white listening subject}, their language is still marked as racialized. Likewise, we find that  writing adapted to be grammatically ``correct" can still be marked as ``non-native".

As discussed in Section \ref{ref:signschange}, the indexical relation that linked an author of a certain language background with a certain quality of science, however, remained unchanged. As a result, though interviewees described many reasons for using ChatGPT to ``polish” their own writing prior to submission, strategies for use often included steps to avoid perceived reviewer indexing of ChatGPT use with language background. In the short term, interviewees described that the availability of ChatGPT made it harder for a reviewer to ascertain the language background of an author; this may explain the small decrease in critiques of writing and the rise in scores we observe in our statistical analysis of reviews. As several interviewees indicated, their ability to guess the identity of an author based on writing became inaccurate at best. 

Yet interviewees found other signs of author identity. For example, the ``amount of experimental work” and ``technical dotting” of a manuscript, as well as a manuscript that was less ``intellectually interesting,”  indicated to one reader that a manuscript would have a long author list, which that reader associated with ``the sheer volume” that ``junior researchers in China” are expected to produce. Another interviewee shared the perception that in non-anonymous venues, even acknowledgment sections that include traditionally Latine names could evoke differential judgment from reviewers: ``If you have, ``we thank John Smith for the English editing of this article” [the reviewers] wouldn't complain about the English at all. But if you have... ``his name was Juan something"... even if [the English] was fine they will judge you differently.” Whether or not this perception is an accurate reflection of reviewer behavior, the anecdote itself influenced how this scholar, themself Latine, presented their own work. 

Because the indexical relationship we observe is not directly between the grammatical errors and the quality of the paper, but grammatical errors with a person and that person with the quality of the paper, the removal of the grammatical errors by ChatGPT does not immediately appear to eliminate bias or the manifestation of bias in scientific peer review. However, the tool's use may dampen the ability for people to detect whether writers were from certain groups. The importance of an intervention against the expression of bias should not be discounted as a valuable contribution to fair scientific review. At the same time, such an intervention may mask shifting ideological work happening on the ground, and only strengthen the symbolic interpretation of language diversity as a deficit to be fixed.

\section{Limitations and Future Work}
There are several limitations to our study which we hope will inspire future work. First, we examine a single conference. Future studies might explore how scientists in other conferences and disciplines engage with language and identity. Second, there may be other confounders which entangle with the ideological work we document. For example, a decrease in writing critiques could be based on perception of LLM-generated writing, as we postulate, but it could also be related to the reviews themselves being generated. Finally, future studies might examine potential interventions to disrupt how reviewers index associations between scholar origin and scholar science quality.  Experimental studies might explore whether this phenomenon is related to the speed at which computer science publishing takes place, and test ways to encourage reviewers to slow down. 
\section{Conclusion}
This article describes the expression and interpretation of language ideologies in a scientific publishing community. We offer multiple contributions. Empirically, we give evidence of how language ideologies shift in response to new technologies. Theoretically, we knit together literature from several disciplinary traditions to understand why this change may occur. 

We find that reviewers critique paper clarity significantly more when authors are affiliated with institutions in countries where English is less widely spoken. We see small shifts in review scores and writing critique after the release of ChatGPT. Interviews with scholars who submitted and reviewed for ICLR in 2023 and 2024 indicate that textual features of ChatGPT can supplement grammatical idiosyncrasies as signs that writing is from a author whose first language is not English. The availability of ChatGPT may mask some of the signs readers associate with writer language background, but readers interpret new signs of difference in anonymized submissions. Interviewees describe that they associate these signs with scholar origin, and scholar origin with the quality of science described in a manuscript. By mapping this chain of signs, we hope to offer room for future scholars to understand the development of biases and intervene against discrimination in scientific publishing.

To this end, we echo Lepp and Sarin \cite{lepp2024global} in recommending more options for multilingual publishing, conferences conducted with local scientists and in local languages, and language coursework requirements for graduate programs. Our findings indicate that instead of considering monolingual publishing as the most efficient path forward, interventions should emphasize the value of different languages and their speakers in the production of science.

\begin{acks}
This study is indebted to the scientists who shared their experiences with us for this project. We dedicate this study to them and to our brilliant academic friends and colleagues from all over the world who make U.S. academia so strong. We thank Neha Nayak Kennard for her  generous support and expertise on OpenReview and on using machine learning to study epistemic values. We are grateful for discussions about research design with Daniel A. McFarland, Angèle Christin, Rebecca Tarlau, Miyako Inoue, Jonathan Rosa, Jennifer Eberhardt, and Dan Jurafsky, as well as the thoughtful manuscript feedback from our anonymous reviewers, Stanford CWIP, and the Coast-to-Coast PhD AI Interest Group. H.L. is supported by a seed grant from the Stanford Institute for Human-Centered Artificial Intelligence
(HAI).
\end{acks}

\bibliographystyle{ACM-Reference-Format}
\bibliography{Supporting_files/main}

%%
%% If your work has an appendix, this is the place to put it.
\appendix
\section{Regression variables}
\label{appendix:variables}
In this section, we discuss the methods for producing our regression variables. 
\subsection{Variables of interest}
We label the unique reviews in the corpus based on their evaluation of writing. Reviewers are anonymous, so there is no way to count unique reviewers. Approximately 48,601 authors are in the dataset; if they created multiple OpenReview accounts, this number could be a low estimate. However, OpenReview makes efforts to merge old and new accounts, so this effect should be minimal. 

\subsubsection{Aspect of evaluation}
Every submission receives a textual review and numerical score score by each reviewer. The score review number options change from year to year, so to measure change over time we create a normalized score by dividing the given score by the maximum possible score for that year. This normalized score sees a large dip in 2024.  

To tease apart the meaning of the score, and how much of the score specifically criticizes writing, we analyze the text of each review. We label every sentence in every review of every paper according to a taxonomy of 6 dimensions outlined in \citep{hua2019argument, kennard2021disapere}. These dimensions, created from a manually annotated set of ICLR reviews, include different aspects of evaluation (e.g. clarity, replicability, thoroughness) as well as the polarity of aspects (critique or praise), making 12 value judgments and 1 residual "none" category. To do this, annotators, ``graduate students in computer science who have undergone training and calibration", first labeled if a given review sentence was evaluative; if so, then whether it was praise or criticism, and then on which criterion. No evaluative sentence can be labeled with multiple criteria. For our dependent variable, clarity critique, annotators were prompted to indicate whether a given critical review sentence expressed concern with the following definition: ``Is the paper clear, well-written, and well-structured?"

We use this method to also create control variables for other epistemic value judgments in each review: critique and praise of accuracy, novelty, thoroughness, consistency, and replicability.

\subsubsection{Author region}
For every author associated with each paper, we extract all self-reported institutional affiliations. Using the Python package \textit{\href{https://github.com/zhijing-jin/email2country?tab=readme-ov-file}{email2country}}, we match the earliest listed institutional domain for each author with a country label, based on literature that indicates it is more likely that the direction of migration throughout a STEM career is from a country of origin to employment or a higher degree or employment in a new region \citep{hanson2016high, braindrain}. We supplement this method by manually labeling countries for unrecognized institutions. Many emails contain non-academic company domains. We manually label each of these companies by country of headquarters unless a country code (e.g. .cn) is included in the domain. We do not give a country code to gmail, outlook, hotmail, or googleemail domains. After 2023, there is an small emergence of domains associated with Ascension Island (.ai). We manually correct typos (such as .eud after university names). We map country to world region and TOEFL requirement as shown below.

\subsubsection{Post-GPT year}
This binary variable does not represent that a specific review is evaluating a paper that was generated by AI. We start with the estimate, based on the work of \cite{liang2024monitoring}, that after 2023 approximately 20\% of submissions contain substantially generated text.

By examining not the specific papers which have generated text, but papers at the year level in which some proportion of the papers use GPT, we can capture two phenomena. First, the variable will pick up change associated with submissions which include generated text. Second, the variable will pick up on indirect behavioral change associated with the refractions of such text being in the publishing ecosystem 
\citep{levy2018privacy}. 
\onecolumn
\begin{table}
\caption{Table of Country, Region, and TOEFL}
\begin{tabular}{lll}
\toprule
       Country & Region &    TOEFL requirement\\
\midrule
                 United States &          North America &   Exempt \\
                     Australia &                Oceania &   Exempt \\
                         China &                   Asia & Required \\
                         India &                   Asia & Required \\
                     Singapore &                   Asia &   Exempt \\
                       Finland &                 Europe & Required \\
                United Kingdom &                 Europe &   Exempt \\
                        Canada &          North America &   Exempt \\
                   Switzerland &                 Europe & Required \\
                       Belgium &                 Europe & Required \\
                   South Korea &                   Asia & Required \\
                         Qatar &                   Asia & Required \\
                       Austria &                 Europe & Required \\
                     Argentina &          South America & Required \\
                        France &                 Europe & Required \\
                        Israel &                   Asia & Required \\
                   New Zealand &                Oceania &   Exempt \\
                   Netherlands &                 Europe & Required \\
                         Italy &                 Europe & Required \\
                         Japan &                   Asia & Required \\
                       Germany &                 Europe & Required \\
                        Brazil &          South America & Required \\
                          Iran &                   Asia & Required \\
                    Bangladesh &                   Asia & Required \\
                     Hong Kong &                   Asia & Required \\
                     Sri Lanka &                   Asia & Required \\
                      Thailand &                   Asia & Required \\
            Russian Federation &                 Europe & Required \\
                        Sweden &                 Europe & Required \\
                        Greece &                 Europe & Required \\
                         Spain &                 Europe & Required \\
                        Turkey &                   Asia & Required \\
                        Taiwan &                   Asia & Required \\
British Indian Ocean Territory &                   Asia & Required \\
                  Saudi Arabia &                   Asia & Required \\
              Ascension Island &                     AI & Required \\
               The Netherlands &                 Europe & Required \\
                       Vietnam &                   Asia & Required \\
          United Arab Emirates &                   Asia & Required \\
                       Ireland &                 Europe &   Exempt \\
                      Slovakia &                 Europe & Required \\
                  South Africa &                 Africa & Required \\
                       Romania &                 Europe & Required \\
                       Lebanon &                   Asia & Required \\
                        Norway &                 Europe & Required \\
                       Algeria &                 Africa & Required \\
                        Mexico &          North America & Required \\
                        Serbia &                 Europe & Required \\
                        Poland &                 Europe & Required \\
                       Croatia &                 Europe & Required \\
                      Colombia &          South America & Required \\
                         Macao &                   Asia & Required \\
                      Pakistan &                   Asia & Required \\
                         Egypt &                 Africa & Required \\
\bottomrule
\end{tabular}
\end{table}

\begin{table}
\begin{tabular}{lll}
\toprule
       Country & Region &    TOEFL requirement\\
\midrule
                European Union &                 Europe & Required \\
                      Bulgaria &                 Europe & Required \\
                      Ukraine &                 Europe & Required \\
                     Lithuania &                 Europe & Required \\
                       Czechia &                 Europe & Required \\
                      Portugal &                 Europe & Required \\
                         Sudan &                 Africa & Required \\
                          Peru &          South America & Required \\
                       Denmark &                 Europe & Required \\
                     Guatemala &          North America & Required \\
                      Malaysia &                   Asia & Required \\
                       Hungary &                 Europe & Required \\
                      Cameroon &                 Africa & Required \\
                       Uruguay &          South America & Required \\
                       Morocco &                 Africa & Required \\
                         Chile &          South America & Required \\
        Bosnia and Herzegovina &                 Europe & Required \\
                        Jordan &                   Asia & Required \\
                        Rwanda &                 Africa & Required \\
                       Tunisia &                 Africa & Required \\
                       Albania &                 Europe & Required \\
                    Luxembourg &                 Europe & Required \\
                       Armenia &                   Asia & Required \\
                       Iceland &                 Europe & Required \\
                     Indonesia &                   Asia & Required \\
                        Russia &                 Europe & Required \\
                      Paraguay &          South America & Required \\
                      Ethiopia &                 Africa & Required \\
                       Estonia &                 Europe & Required \\
                        Kuwait &                   Asia & Required \\
                          Oman &                   Asia & Required \\
                    Madagascar &                 Africa & Required \\
                    Kazakhstan &                   Asia & Required \\
                         Nepal &                   Asia & Required \\
                   Philippines &                   Asia & Required \\
                       Georgia &                   Asia & Required \\
                      Slovenia &                 Europe & Required \\
                         Kenya &                 Africa & Required \\
                        Kosovo &                 Europe & Required \\
                         Ghana &                 Africa & Required \\
                        Latvia &                 Europe & Required \\
                       Belarus &                 Europe & Required \\
                       Nigeria &                 Africa & Required \\
                        Uganda &                 Africa & Required \\
                        Cyprus &                   Asia & Required \\
                    Montenegro &                 Europe & Required \\
                    Cape Verde &                 Africa & Required \\
          Syrian Arab Republic &                   Asia & Required \\
                     Venezuela &          South America & Required \\
                         Yemen &                   Asia & Required \\
               North Macedonia &                 Europe & Required \\
                          Cuba &          North America & Required \\
                    Costa Rica &          North America & Required \\
       Cocos (Keeling) Islands &                   Asia & Required \\
\bottomrule
\end{tabular}
\end{table}

\subsection{Descriptive analysis of evaluative categories}
\begin{figure}[!th]
     \centering
    \raisebox{-\height}{\includegraphics[width=\textwidth]{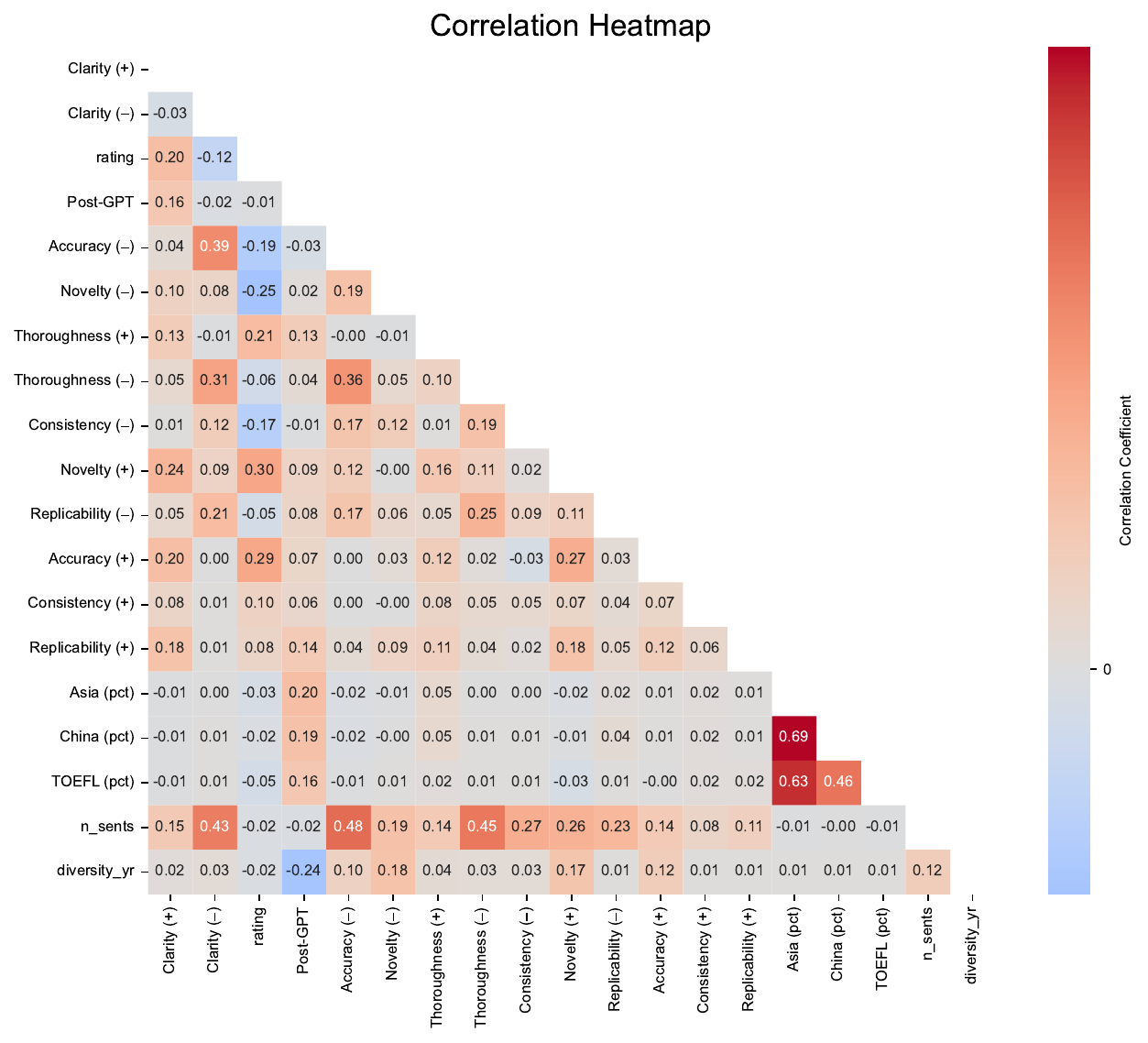}}
    \caption{Correlation heatmap of evaluative categories}
    \Description[Heatmap of evaluative categories] {Heatmap of evaluative categories} 
\end{figure}
\clearpage
\begin{table}[h!]
\caption{Regression variables}
\begin{tabular}{lrrrrrrrr}
\toprule
 & count & mean & std & min & 25\% & 50\% & 75\% & max \\
\midrule
Clarity (+) & 76453.00 & 0.64 & 0.86 & 0.00 & 0.00 & 0.00 & 1.00 & 12.00 \\
Clarity (–) & 76453.00 & 2.14 & 2.89 & 0.00 & 0.00 & 1.00 & 3.00 & 48.00 \\
rating & 76453.00 & 5.27 & 1.75 & 1.00 & 4.00 & 5.00 & 6.00 & 10.00 \\
Post-GPT & 76453.00 & 0.55 & 0.50 & 0.00 & 0.00 & 1.00 & 1.00 & 1.00 \\
Accuracy (–) & 76453.00 & 2.50 & 2.50 & 0.00 & 1.00 & 2.00 & 4.00 & 34.00 \\
Novelty (–) & 76453.00 & 1.26 & 1.55 & 0.00 & 0.00 & 1.00 & 2.00 & 17.00 \\
Thoroughness (+) & 76453.00 & 0.68 & 0.92 & 0.00 & 0.00 & 0.00 & 1.00 & 11.00 \\
Thoroughness (–) & 76453.00 & 3.50 & 3.06 & 0.00 & 1.00 & 3.00 & 5.00 & 51.00 \\
Consistency (–) & 76453.00 & 0.88 & 1.29 & 0.00 & 0.00 & 0.00 & 1.00 & 19.00 \\
Novelty (+) & 76453.00 & 1.57 & 1.60 & 0.00 & 0.00 & 1.00 & 2.00 & 14.00 \\
Replicability (–) & 76453.00 & 0.40 & 0.82 & 0.00 & 0.00 & 0.00 & 1.00 & 15.00 \\
Accuracy (+) & 76453.00 & 0.77 & 1.02 & 0.00 & 0.00 & 0.00 & 1.00 & 9.00 \\
Consistency (+) & 76453.00 & 0.17 & 0.44 & 0.00 & 0.00 & 0.00 & 0.00 & 5.00 \\
Replicability (+) & 76453.00 & 0.08 & 0.31 & 0.00 & 0.00 & 0.00 & 0.00 & 7.00 \\
Africa (pct) & 76453.00 & 0.38 & 4.87 & 0.00 & 0.00 & 0.00 & 0.00 & 100.00 \\
Asia (pct) & 76453.00 & 37.12 & 37.80 & 0.00 & 0.00 & 25.00 & 71.43 & 100.00 \\
Europe (pct) & 76453.00 & 19.44 & 33.14 & 0.00 & 0.00 & 0.00 & 25.00 & 100.00 \\
N. America (pct) & 76453.00 & 40.23 & 37.33 & 0.00 & 0.00 & 33.33 & 75.00 & 100.00 \\
Oceania (pct) & 76453.00 & 1.70 & 9.51 & 0.00 & 0.00 & 0.00 & 0.00 & 100.00 \\
S. America (pct) & 76453.00 & 0.35 & 4.05 & 0.00 & 0.00 & 0.00 & 0.00 & 100.00 \\
n sentences & 76453.00 & 29.06 & 20.89 & 1.00 & 18.00 & 25.00 & 35.00 & 621.00 \\
Shannon diversity by year & 76453.00 & 2.39 & 0.05 & 2.20 & 2.38 & 2.40 & 2.40 & 2.47 \\
Reviewers (N) & 20827.00 & 3.67 & 0.64 & 1.00 & 3.00 & 4.00 & 4.00 & 9.00 \\
Authors (N) & 17448.00 & 2.79 & 1.80 & 1.00 & 1.00 & 2.00 & 4.00 & 34.00 \\
\bottomrule
\end{tabular}
\end{table}
\clearpage
\begin{table}[h!]
\caption{Model Comparison for effect of region on praise of writing clarity before and after ChatGPT availability}
\begin{center}
\begin{tabular}{lcccccc}
\toprule
\textbf{Dep. Variable}         &   \multicolumn{6}{c}{Clarity (+) }      \\
\textbf{ChatGPT available}         &   \multicolumn{3}{c}{Before} & \multicolumn{3}{c}{After} \\
\textbf{Estimator}             &  \multicolumn{6}{c}{RandomEffects}          \\
\textbf{No. Observations}      &      34066       &      34066       &      34066       &      42387       &      42387       &      42387        \\
\textbf{Cov. Est.}             &   \multicolumn{6}{c}{Clustered}          \\
\textbf{R-squared}             &      0.0706      &      0.0705      &      0.0700      &      0.1800      &      0.1805      &      0.1797       \\
\textbf{R-Squared (Within)}    &      0.0442      &      0.0442      &      0.0442      &      0.0860      &      0.0860      &      0.0860       \\
\textbf{R-Squared (Between)}   &      0.1283      &      0.1281      &      0.1266      &      0.3600      &      0.3611      &      0.3593       \\
\textbf{R-Squared (Overall)}   &      0.0711      &      0.0711      &      0.0706      &      0.1879      &      0.1884      &      0.1877       \\
\textbf{F-statistic}           &      184.68      &      184.51      &      183.19      &      664.43      &      666.73      &      663.14       \\
\textbf{P-value (F-stat)}      &      0.0000      &      0.0000      &      0.0000      &      0.0000      &      0.0000      &      0.0000       \\
\midrule
\textbf{Asia (pct)}            &     -0.0285      &                  &                  &     -0.0183      &                  &                   \\
\textbf{ }                     &    (-5.8599)     &                  &                  &    (-3.7400)     &                  &                   \\
\textbf{China (pct)}           &                  &     -0.0306      &                  &                  &     -0.0271      &                   \\
\textbf{ }                     &                  &    (-5.7888)     &                  &                  &    (-6.0296)     &                   \\
\textbf{TOEFL (pct)}           &                  &                  &     -0.0178      &                  &                  &     -0.0081       \\
\textbf{ }                     &                  &                  &    (-3.8702)     &                  &                  &    (-1.5529)      \\
\bottomrule
\end{tabular}
\end{center}
\end{table}
\begin{table}[h!]
\caption{Model Comparison for effect of region on critique of writing clarity before and after ChatGPT availability}
\begin{center}
\begin{tabular}{lcccccc}
\toprule
\textbf{Dep. Variable}         &   \multicolumn{6}{c}{Clarity (-) }      \\
\textbf{ChatGPT available}         &   \multicolumn{3}{c}{Before} & \multicolumn{3}{c}{After} \\
\textbf{Estimator}             &  \multicolumn{6}{c}{RandomEffects}          \\
\textbf{No. Observations}      &      34066       &      34066       &      34066       &      42387       &      42387       &      42387        \\
\textbf{Cov. Est.}             &   \multicolumn{6}{c}{Clustered}          \\

\textbf{R-squared}             &      0.2435      &      0.2433      &      0.2434      &      0.3967      &      0.3967      &      0.3968       \\
\textbf{R-Squared (Within)}    &      0.2494      &      0.2494      &      0.2494      &      0.3893      &      0.3893      &      0.3893       \\
\textbf{R-Squared (Between)}   &      0.2288      &      0.2280      &      0.2284      &      0.4156      &      0.4157      &      0.4159       \\
\textbf{R-Squared (Overall)}   &      0.2420      &      0.2417      &      0.2419      &      0.3986      &      0.3987      &      0.3987       \\
\textbf{F-statistic}           &      782.74      &      781.95      &      782.50      &      1990.0      &      1990.2      &      1990.7       \\
\textbf{P-value (F-stat)}      &      0.0000      &      0.0000      &      0.0000      &      0.0000      &      0.0000      &      0.0000       \\
\midrule
\textbf{Asia (pct)}            &      0.0255      &                  &                  &      0.0049      &                  &                   \\
\textbf{ }                     &     (4.1238)     &                  &                  &     (1.1538)     &                  &                   \\
\textbf{China (pct)}           &                  &      0.0211      &                  &                  &      0.0069      &                   \\
\textbf{ }                     &                  &     (3.0512)     &                  &                  &     (1.7215)     &                   \\
\textbf{TOEFL (pct)}           &                  &                  &      0.0219      &                  &                  &      0.0112       \\
\textbf{ }                     &                  &                  &     (3.9241)     &                  &                  &     (2.5287)      \\
\bottomrule
\end{tabular}
\end{center}
\end{table}
\begin{table}[h!]
\caption{Model Comparison for effect of region on reviewer rating before and after ChatGPT availability}
\begin{center}
\begin{tabular}{lcccccc}
\toprule
\textbf{Dep. Variable}         &   \multicolumn{6}{c}{Reviewer Rating}      \\
\textbf{ChatGPT available}         &   \multicolumn{3}{c}{Before} & \multicolumn{3}{c}{After} \\
\textbf{Estimator}             &  \multicolumn{6}{c}{RandomEffects}          \\
\textbf{No. Observations}      &      34066       &      34066       &      34066       &      42387       &      42387       &      42387        \\
\textbf{Cov. Est.}             &   \multicolumn{6}{c}{Clustered}          \\
\textbf{R-squared}             &      0.2734      &      0.2730      &      0.2737      &      0.2598      &      0.2596      &      0.2600       \\
\textbf{R-Squared (Within)}    &      0.2303      &      0.2303      &      0.2304      &      0.2177      &      0.2177      &      0.2177       \\
\textbf{R-Squared (Between)}   &      0.3505      &      0.3496      &      0.3514      &      0.3504      &      0.3501      &      0.3511       \\
\textbf{R-Squared (Overall)}   &      0.3068      &      0.3063      &      0.3072      &      0.2878      &      0.2875      &      0.2882       \\
\textbf{F-statistic}           &      853.95      &      852.45      &      855.42      &      991.36      &      990.64      &      992.69       \\
\textbf{P-value (F-stat)}      &      0.0000      &      0.0000      &      0.0000      &      0.0000      &      0.0000      &      0.0000       \\
\midrule
\textbf{Asia (pct)}            &     -0.0381      &                  &                  &     -0.0145      &                  &                   \\
\textbf{ }                     &    (-5.6062)     &                  &                  &    (-2.5550)     &                  &                   \\
\textbf{China (pct)}           &                  &     -0.0312      &                  &                  &     -0.0026      &                   \\
\textbf{ }                     &                  &    (-4.1587)     &                  &                  &    (-0.4926)     &                   \\
\textbf{TOEFL (pct)}           &                  &                  &     -0.0423      &                  &                  &     -0.0260       \\
\textbf{ }                     &                  &                  &    (-6.6278)     &                  &                  &    (-4.3805)      \\
\bottomrule

\end{tabular}
\end{center}
\end{table}
\clearpage

\section{Interview Protocols}
\begin{samepage}
\subsection{Author focused}
\begin{obeylines}
Can you tell me a bit about your life and what brought you to become an AI scientist?
What is your title?  
Do you identify with a country of origin? 
Where do you work now?
What languages or language varieties do you speak?
Where did you learn these languages?
In what ways do you use that language in your work? At home? With friends? Family? 
Do you work mostly with people who speak the same language(s) as you, or people who speak different language(s)/language varieties?
What has been your experience with academic writing in language other than English?
Where do you find new research to read? Do you read in paper form or other venues? What languages are they generally written in? Have you ever written or considered writing in your other language(s)? Why do you think this is?
What has been your experience with peer review? Have you ever peer reviewed?
Have you ever come across papers as a reviewer in which you think the writing or language is not appropriate for publication? What are the elements of the paper that make you think that? What do you do as a reviewer when this happens?
Have you ever had a reviewer talk about the language of a paper you wrote? If so, what did you do in response?
What is your writing process like when submitting a paper to a conference such as ICLR? 
What platform do you use to write? Do you use any text-editing integrations? What are the biggest benefits to using these tools? Are there any drawbacks to using these tools?
Do you ever disagree with what the person or technological tool suggests? Do you accept or decline their changes? How much do you think these tools change your original writing? What kinds of changes do you notice?
Why do you use these tools? 
If you were to write a paper for an audience of people who speak <otherlanguage>, how similar/different would it look from a paper that you submit to ICLR? 
What do you think are the biggest challenges to increasing international diversity in computer science publishing? 
Is there anything else you want to tell me about? Do you know anybody else I might talk to?
\end{obeylines}
\end{samepage}

\subsection{AC/Reviewer focused}
\begin{obeylines}
Today, we’re going to talk about what it’s like to write and review for ICLR. If there are any questions you don’t want to answer, please feel free to skip them.
Can you tell me a bit about your life and what brought you to become an AI scientist/AC for ICLR?
What is your title?  
Do you identify with a country of origin? 
Where do you work now?
What languages or language varieties do you speak?
Where did you learn these languages?
In what ways do you use that language in your work? At home? With friends? Family? 
Do you work mostly with people who speak the same language(s) as you, or people who speak different language(s)/language varieties?
What has been your experience with reviewing? How has it changed over the last several years?
How do you perceive international diversity at ICLR? Where are people from? 
What is your working process like when evaluating papers and reviews at a conference such as ICLR? 
What are the qualities of reviews/papers that you value?
Have you ever disagreed with reviewers? What do you do in that case?
Have you ever come across papers in which you think the writing or language is not appropriate for publication? What are the elements of the paper that make you think that? What do you do as a reviewer when this happens?
Have you noticed a change in writing since GPT was released?
What do you think are the biggest challenges to increasing international diversity in computer science publishing? 

\end{obeylines}
\section{Interview Codes}
Indexing: Top level code to describe indexical relationships
\begin{enumerate}
    \item \textbf{Detected accent with some quality of science} Science may be described as less useful; fewer resources; more junior. Logic that fixing accent fixes science.
    \item \textbf{Detected accent with person country identity} May describe native/non-native speakers, people from England or the U.S. vs. people from other countries
   \item \textbf{Avoiding perceived indexical relationships}	Describing that one must avoid sounding like AI, must remove accent
    \item \textbf{Person country identity with some quality of science} Person may be from other country, especially U.S. or China. Science may be described as less useful; fewer resources; more junior or positively.
    \item \textbf{China-specific} When this code is used to describe people from China specifically
    \item \textbf{GPT with some quality science} Certain words described as being associated with GPT tied to science quality. Science may be described as less useful; fewer resources; more junior.
    \item \textbf{GPT with person country identity} Certain words described as being associated with \item \textbf{GPT tied to person country identity or other association with country} For example, high author count, different style of writing, etc. associated with being from certain country.
\end{enumerate}
Language labor: Top level code to describe labor to accommodate English-only publishing.
\begin{enumerate}
\item \textbf{Languages} Languages spoken
\item \textbf{School} Language challenges in academic life 
\item \textbf{Country transition}	Narratives about moving for academic or job purposes
\item \textbf{Graduate}	Narratives about graduate school
\item \textbf{Undergraduate}	Narratives about undergraduate school
\item \textbf{Why English}	Narratives about studying English
\item \textbf{Why move}	Narratives about purpose of immigration
\item \textbf{Social network}	Working with others, academically or socially
\item \textbf{Collaborators}	Writing together with collaborators
\item \textbf{Speaking and listening}	Speaking and listening experiences in scientific community
\item \textbf{Reading} Challenges or success related to reading in order to be a part of a scientific community
\item \textbf{Writing}	Writing together with GPT
\item \textbf{GPT Prompt}	How people prompt GPT
    \item \textbf{GPT Use}	How people use GPT
    \item \textbf{Grammarly}	How people use Grammarly
\item \textbf{Non-academic writing}	Writing not for publication, such as for class or for family
\item \textbf{Non-English writing}	Writing about work in other languages
\item \textbf{Receiving reviews}	Experiences with receiving reviews
\item \textbf{Reviewing} Challenges or success related to reviewing in order to be a part of a scientific community
\item \textbf{Anonymity} Trying to break anonymity standards or perceiving others to
\item \textbf{Guessing L1} Guessing at first language of person based on writing
\item \textbf{Translation}	Uses of translation
\end{enumerate}

\end{document}